\documentstyle[version2,aps]{revtex} 
\begin{document}                                                        
\draft                                                                  
\twocolumn
\widetext                                                             
\begin{title}
Spin dynamics of the doped t-J model
\end{title}

\author{R. Eder and Y. Ohta}

\begin{instit}
Department of Applied Physics, Nagoya University, Nagoya 464-01, Japan
\end{instit}

\begin{abstract}
Using the exact diagonalization technique we study the low energy 
spin excitations in moderately doped (hole concentration $<25$\%)
finite clusters of $t-J$ model. To clarify whether a given low energy 
spin excitation corresponds to a particle-hole transition in the
`quasiparticle band' near the Fermi energy, we compare the electron 
addition and removal spectra of the respective final state 
to that of the ground state. We find that the low energy
spin excitation spectrum 
is composed of different branches, with the dominant low-energy spin 
excitation at $(\pi,\pi)$ being a spin-wave like collective mode,
whereas the low energy spin excitations with momentum transfer
other than $(\pi,\pi)$ correspond to particle-hole excitations.
\end{abstract} 

\pacs{pacs 74.20.-Z, 75.10.Jm, 75.50.Ee}
\narrowtext
\topskip8cm
The spin dynamics of strongly correlated electron systems is
a key issue for the understanding of high-temperature
superconductors. As an experimental fact, neutron scattering experiments 
show a broad peak in the dynamical spin correlation function
for in-plane momentum transfer $(\pi,\pi)$ at low
frequencies $\sim 40 meV$\cite{RossatMignod}. Exact diagonalization 
studies of small clusters of $t-J$ model usually reproduce 
this behaviour\cite{Dagottoetal,TohyamaMaekawa}, 
and it is the purpose of the present work to distinguish between
different interpretations which are possible for this peak:
within a weak coupling-like picture, the dynamical 
spin susceptibility is determined by particle-hole excitations
of either real electrons or `spinons', the strong scattering intensity
for momentum transfer $(\pi,\pi)$ then would be the consequence
of Fermi surface nesting. A different interpretation would be to 
attribute the peak to the remnants of the short wavelength
spin waves present in the undoped compounds; as long as their wavelength
is shorter than the spin correlation length, such spin wave like
excitations might well persist, but being no longer the Goldstone modes 
associated with antiferromagnetically broken symmetry, a finite 
frequency would be naturally expected\cite{KuchievSushkov}. 
Our results show that in finite $t-J$ model 
clusters (lattice size $\leq20$) with low hole concentration
the second interpretation is the correct one: 
the dominant low-energy peak at $(\pi,\pi)$ corresponds to a collective
spin wave-like mode, particle hole excitations are restricted to 
different momentum transfer. The $t$$-$$J$ model reads:
\[
H =                                                    
 -t \sum_{< i,j >, \sigma}                                         
( \hat{c}_{i, \sigma}^\dagger \hat{c}_{j, \sigma}  +  H.c. )
 + J \sum_{< i,j >}(\bbox{S}_i \cdot
 \bbox{S}_j
 - \frac{n_i n_j}{4}).
\]
Here the $\bbox{S_i}$ are the electronic 
spin operators and                                                             
the sum over $<i,j>$ stands for a summation                                    
over all pairs of nearest neighbors.
The operators $\hat{c}_{i,\sigma}$
are expressed in terms of ordinary fermion                                  
operators as $c_{i,\sigma}(1-n_{i,-\sigma})$.\\
To clarify the nature of the low-energy spin excitations, we
applied the following scheme: using the Lanczos algorithm
we computed the dynamical spin correlation function (SCF)
\begin{eqnarray}
S(\bbox{q},\omega) &=&
\sum_{\nu} | \langle \Psi_{\nu,n} | S^{z}(\bbox{q})|
\Psi_{0,n} \rangle |^2
\nonumber \\
&\;&\;\;\;\delta( \omega - ( E_{\nu,n} - E_{0,n} )),
\end{eqnarray}
\topskip0cm
where $|\Psi_{\nu,n} \rangle$  ($E_{\nu,n}$)  
is the $\nu^{th}$ eigenstate (eigenenergy) with
$n$ holes (in particular $\nu$$=$$0$ implies the ground state).\\
Fig. \ref{fig1} shows $S(\bbox{q},\omega)$
in the $18$-site cluster 
with $2$ and the $16$-site cluster with $4$ holes
(in this paper we restrict ourselves to these systems, whose 
behaviour is representative). We then extract the energies 
$E_{\nu,n}$ of the dominant low-energy peaks
indicated by arrows in Fig. \ref{fig1}, these are the
two lowermost peaks for each cluster.
Each of them is a single, unsplit peak
so that a unique (and for the low energy peaks highly precise)
value for $E_{\nu,n}$ can be obtained.
Using the inverse iteration algorithm\cite{Jennings} 
we can now converge out the corresponding final state wave 
function $|\Psi_{\nu,n}\rangle$
(using a reasonably accurate estimate $E_{tr}$ for an eigenvalue, 
the inverse iteration method essentially consists in applying powers 
of $(H - E_{tr})^{-1}$ to some trial state).
Thereby we also obtain an independent estimate for
$E_{\nu,n}$, which always agreed with the value extracted
from the correlation function 
to an accuracy better than $10^{-10}$, essentially the 
limit of the Lanczos procedure; 
total spin and point group symmetry of the obtained
$|\Phi_{\nu}\rangle$ were cross-checked to be compatible with the 
selection rules. In the last step we calculated
(again via the Lanczos method) the electronic spectral functions 
$A^{(\alpha)}
(\bbox{k}, \omega)=
A^{(\alpha)}_{n,-}(\bbox{k},-\omega) +
A^{(\alpha)}_{n,+}(\bbox{k}, \omega)$, where
\begin{eqnarray}
A^{(\alpha)}_{n,-}(\bbox{k}, \omega) &=&
\sum_{\mu} | \langle \Psi_{\mu,n+1} | \hat{c}_{\bbox{k},\sigma}|
\Psi_{\alpha} \rangle |^2
\nonumber \\
&\;&\;\;\;\delta( \omega - ( E_{\mu,n+1} - E_{0,n} )),
\nonumber \\
A^{(\alpha)}_{n,+} (\bbox{k}, \omega) &=&
\sum_{\mu} | \langle \Psi_{\mu,n-1} 
| \hat{c}_{\bbox{k},\sigma}^\dagger |
\Psi_{\alpha} \rangle |^2
\nonumber \\
&\;&\;\;\;\delta( \omega - ( E_{\mu,n-1} - E_{0,n} )).
\label{spectral}
\end{eqnarray}
\topskip0cm
where
$|\Psi_{\alpha=0} \rangle$ denotes the $n$-hole ground state
and $|\Psi_{\alpha=1} \rangle$ the SCF-final state.
These functions are simply the photoemission (PES) and inverse
photoemission (IPES) spectra for the ground state and 
SCF-final state. By inspection of the difference
$A_d(\bbox{k},\omega) = A^{(1)}(\bbox{k},\omega) -
A^{(0)}(\bbox{k},\omega)$
(which henceforth will be referred to as the 
`difference spectrum for the state $|\Psi_{\nu,n}\rangle$')
we can now decide whether the creation of a
spin excitation is associated with the transfer of single particle
spectral weight in the band near the Fermi energy or not.\\ 
To begin with, Fig. \ref{fig2}
shows the single particle spectral function
of the ground state and the difference spectrum
for the SCF final state for momentum transfer
$\bbox{q}=(2\pi/3,2\pi/3)$.
In order to eliminate shifts of spectral weight on
small energy scales, a relatively large 
Lorentzian broadening of $0.4t$ is used. 
We locate the Fermi energy as separating
the highest PES and lowest IPES peaks in the ground state spectra.
The difference spectra have remarkably
small weight as compared to the original spectra, particularly
the parts remote from $E_F$: small `wiggles',
indicate slight shifts of `peaks' within the PES spectrum. Major 
changes of the spectral function are restricted to the
neighborhood of $E_F$: at $\bbox{k}_1=(-2\pi/3,0)$, 
PES weight is shifted to the IPES side right at $E_F$,
indicating that at this momentum the SCF-final state has a reduced 
electron occupancy of the quasiparticle band
as compared to the ground state.
At $\bbox{k}_1+\bbox{q}=(0,2\pi/3)$ the reverse can be 
seen, i.e. the SCF-final state has an enhanced electron occupancy
of the quasiparticle band. We thus see a shift of low energy
spectral weight with momentum transfer $\bbox{q}$, precisely 
the expected signature of a particle-hole excitation.
To a lesser degree, such a shift can also be seen between
$(\pi/3,-\pi/3)$ and $(\pi,\pi/3)$, but the weight transfer
is substantially smaller.\\
The particle-hole character is even more pronounced 
for the low energy spin excitation with momentum transfer
$\bbox{q}=(\pi/2,\pi/2)$ in the $16$-site cluster with 
$4$ holes (Fig. \ref{fig3}) where 
we clearly see the transfer of an electron right at $E_F$
from $(\pi,0)$ to $(\pi/2,-\pi/2)$. We conclude that
our technique indeed provides a sensitive tool for
detecting particle hole excitations and consequently apply it to
the dominant low energy peak at $(\pi,\pi)$.
The difference spectra for the respective final states in the
$18$ and $16$-site clusters are shown in Figs. \ref{fig4} and
\ref{fig5}. There are `wiggles' near the Fermi energies  in some of 
them which indicate 
a shift of the quasiparticle peaks. There is, however, 
no appreciable shift of weight from PES to IPES near the
Fermi energy, despite the fact that the spectral intensity of
these excitations in $S(\bbox{q},\omega)$ is much stronger
than the particle-hole excitations considered previously.
The spin excitation at $(\pi,\pi)$ thus obviousy is not associated
with a shift of spectral weight near the Fermi energy, i.e. it 
is not a particle-hole excitation but rather a collective mode
the character of which remains to be clarified.\\
Up to a constant, the spin correlation function for 
this momentum equals the correlation function of the
staggered magnetization, $M_S=\sum_{i\in A} S_i^z - \sum_{i\in B} S_i^z$.
We will now assume that the ground state expectation value 
$\langle M_S^2 \rangle =(N\cdot m_S)^2$, 
where $N$ denotes the number of sites and $m_S$ is of order unity,
and show that this allows the explicit construction of a low energy 
collective spin excitation with momentum transfer $(\pi,\pi)$
(we do not assume $\langle M_S \rangle\neq 0$, i.e. there need not be
broken symmetry). 
We decompose the ground state wave function $|\Psi_0 \rangle$ as
$|\Psi_0\rangle = \sum_\nu g_\nu |\psi_\nu\rangle$,
where the $|\psi_\nu\rangle$ are defined by the requirements
$M_S|\psi_\nu\rangle = \nu |\psi_\nu\rangle$ and
$\langle\psi_\nu|\psi_\nu\rangle=1$. In other words, we decompose
the ground state into components with fixed value of
the staggered magnetization. We define the matrix elements
of the Hamiltonian, $h_{\nu,\mu} =
\langle \psi_\nu | H | \psi_\mu \rangle$, and obviously have
$h_{\nu,\mu}=0$ for $|\nu-\mu|>2$ 
(the Hamiltonian can change the staggered magnetization at 
most by $2$). The requirement that the ground state 
energy $E_0$ be extensive then implies $h_{\mu,\nu} \sim N$.\\
Introducing the translation operator by one lattice 
site, $T_{1,0}$, the translational invariance of $|\Psi_0\rangle$ 
guarantees that $g_\nu T_{1,0} |\psi_\nu\rangle = g_{-\nu} 
|\psi_{-\nu}\rangle$. We now define the trial state
$|\Psi_1\rangle = \sum_{\nu>0} (g_\nu |\psi_\nu\rangle -
g_{-\nu} |\psi_{-\nu}\rangle)$. By construction this state changes sign
under $T_{1,0}$, i.e. it has momentum $(\pi,\pi)$.
The expectation values of the Hamiltonian with 
$|\Psi_0\rangle$ and $|\Psi_1\rangle$ differ by
$\Delta H = h_{0,0} g_0^2 + 4 h_{1,0} g_1 g_0 +
4 h_{2,0} g_2 g_0 + + 4  h_{1,-1} g_1 g_{-1}$, the norms of these 
state differ by $\Delta N = g_0^2$. The energy $\Delta =
(\Delta H + E_0 \Delta N)/(1-\Delta N)$ thus  provides a rigorous upper
bound for the energy difference between ground state and
lowest state with total momentum $(\pi,\pi)$ and 
restricting the Hilbert space to $|\Psi_0\rangle$ and $|\Psi_1\rangle$ 
we find $S(\bbox{Q},\omega) \sim m_S^2 \delta(\omega-\Delta)$.\\
Fig. \ref{fig6} shows different forms of the probability distribution
$g_\nu^2$ which are compatible with the requirement that 
$\langle M_S^2 \rangle = \sum_\nu \nu^2 g_\nu^2 \sim N^2$.
Form (a) implies that $g_\nu=0$ for $\nu \sim 0$ so that
$\Delta$$=$$0$. Hence $|\Psi_0\rangle$ and $|\Psi_1\rangle$ are 
degenerate and e.g.  the state $|\Psi_+ \rangle =
\sum_{\nu>0} g_\nu |\psi_\nu\rangle$
by itself is already an eigenstate: this corresponds to true
broken symmetry. Form (b) implies a nonvanishing $\Delta$; here
the number of $g_\nu^2$ which differ from $0$ must be of order $N$, 
so that $g_\nu \sim 1/\sqrt{N}$ and hence $\Delta \sim N^0$:
we have a potentially low energetic collective mode in the spin 
excitation spectrum for all system sizes, which corresponds
to a change of the relative phase of the two components of the ground 
state wave function with positive and negative staggered magnetization.
If $g_\nu^2$ continuously approaches the broken symmetry form (a), 
the excitation energy of this mode will approach zero.
For completeness we note that $\langle M_S^2\rangle \sim N$
(as is the case for a free electron gas) would imply
$\Delta \sim \sqrt{N}$, so that we cannot get a meaningful
low energy mode from the above construction.\\ 
Let us now check this interpretation of the $(\pi,\pi)$ mode.
The exact diagonalization technique gives $|\Psi_0\rangle$
directly as a linear combination of basis states
$\hat{c}_{i_1,\sigma_1}^\dagger \hat{c}_{i_2,\sigma_2}^\dagger 
\dots \hat{c}_{i_n,\sigma_n}^\dagger |vac\rangle$,
so that it is easy to construct 
the state $|\Psi_1\rangle$ (normalized to unity).
We can then compute its overlap with the
SCF final state for momentum transfer $(\pi,\pi)$, $|\Psi_{SCF}\rangle$,
i.e. the quantity $|\langle \Psi_1 | \Psi_{SCF} \rangle|^2$.
Results for various cluster sizes and dopings
are summarized in Tab. \ref{tab1}, which quite obviously confirms
our hypothesis: there is always
a substantial overlap between the respective $|\Psi_1\rangle$
and the exact SCF final state, and for hole concentrations
around $10$\% (2 holes) this overlap is quite comparable to that
in the undoped Heisenberg antiferromagnet, where we can expect
that the $(\pi,\pi)$ mode will evolve into the spin wave mode
for infinite system size.
The finite energy collective mode, which we could predict on quite 
general assumptions thus indeed seems to be
realized in the small clusters. From our general contruction
we expect that this mode persists also in the infinite system
and thus may explain the neutron scattering experiments.\\
In summary, our results show a fairly complicated
picture of the spin excitation spectrum
in the doped two-dimensional $t$$-$$J$ model. 
For low doping levels, (i.e. $2$  or $4$ holes) 
the dominant low energy mode is located at $(\pi,\pi)$, 
and we have identified this mode as a collective mode comparable to
the spin waves present in the undoped cluster. Away from $(\pi,\pi)$ 
on the other hand,
we have fairly conventional particle-hole excitations.\\
The fact that parts of the spin excitation spectrum of the
Heisenberg antiferromagnet persist in the doped system
appears very natural if one adopts the picture that the doped
system should be modelled as an incoherent `spin background'
in which spin-bag-like quasiparticles
corresponding to the doped holes are moving\cite{spinbags}.
Then, it seems natural to distinguish between two types of
spin excitations: the first are the particle-hole excitations
of the hole-liquid, which should resemble that of a Fermi liquid
with a Fermi surface volume proportional to the number of holes. 
This part of the excitation spectrum
can quantitatively explain the Pauli susceptibility at low
temperatures\cite{Trugman}. 
In addition there are the excitations of the
`spin background' which may, with some modifications, resemble
that of the Heisenberg antiferromagnet. 
One example here is the spin wave-like mode at $(\pi,\pi)$,
a possible other example the `two magnon' excitation observed in 
the Raman spectrum until well in the superconducting doping 
regime\cite{Blumberg}.\\
To conclude, we note that the general scheme developed above
allows for a highly selective study of the character of arbitrary
low energy excitations; it thus seems possible 
(at least in principle) to obain
a fairly complete picture of the entire low-energy
excitation spectrum of the 2-d $t-J$ model clusters.\\ 
It is a pleasure for us to acknowledge numerous instructive
discussions with Professor S. Maekawa.
Financial support of R. E. by the Japan Society for the Promotion
of Science is most gratefully acknowledged. Parts of the calculations
were arried out at the Computer Center of the Institute for
Molecular Science, Okazaki.

%%%%%%%%%%%%%%%%%%%%%%%%%%%%%%%%%%%%%%%%%%%%%%%%%%%%%%%%%%%%%%%%%
\figure{Spin correlation function in the $18$-site cluster with
        $2$ holes (top) and the $16$ site cluster with $4$ holes (bottom).
\label{fig1}}

\figure{Difference spectrum for the SCF final state 
        in the $18$-site cluster with $2$ holes
        with momentum $(\frac{2\pi}{3},\frac{2\pi}{3})$.
        The black (gray) area corresponds to IPES
        (PES). Also shown is the
        single particle spectral function for the ground state
        (full line: PES, dotted line: IPES).
        The vertical line marks the Fermi energy. 
\label{fig2}}
\figure{Difference spectrum for the SCF final state 
        in the $16$-site cluster with $4$ holes
        with momentum $(\frac{\pi}{2},\frac{\pi}{2})$.
        All conventions are like in Fig. \ref{fig2}. 
\label{fig3} }
\figure{Difference spectrum for the SCF final state 
        in the $18$-site cluster with $2$ holes
        with momentum $(\pi,\pi)$.
        All conventions are like in Fig. \ref{fig2}. 
\label{fig4} }
\figure{Difference spectrum for the SCF final state 
        in the $16$-site cluster with $4$ holes
        with momentum $(\pi,\pi)$.
        All conventions are like in Fig. \ref{fig2}. 
\label{fig5} }
\figure{Two possible forms of the `wave function' $g_\nu$.
       (a) implies broken symmetry, (b) implies no broken symmetry.
\label{fig6} }
\begin{table}
\caption{Overlap of the exact SCF-final state and the
         trial state based on the spin-wave picture for different
         cluster sizes and dopings. $20$ sites and $4$ holes exceeds
         our computer capacity.}
\begin{tabular}{c|c c c}
$\;$ & 16-site & 18-site & 20-site\\
\tableline
0 holes & 0.8378 & 0.7560 & 0.8214\\
2 holes & 0.7187 & 0.7388 & 0.7059\\
4 holes & 0.4346 & 0.5620 & $\;$  \\
\end{tabular}
\label{tab1}
\end{table}
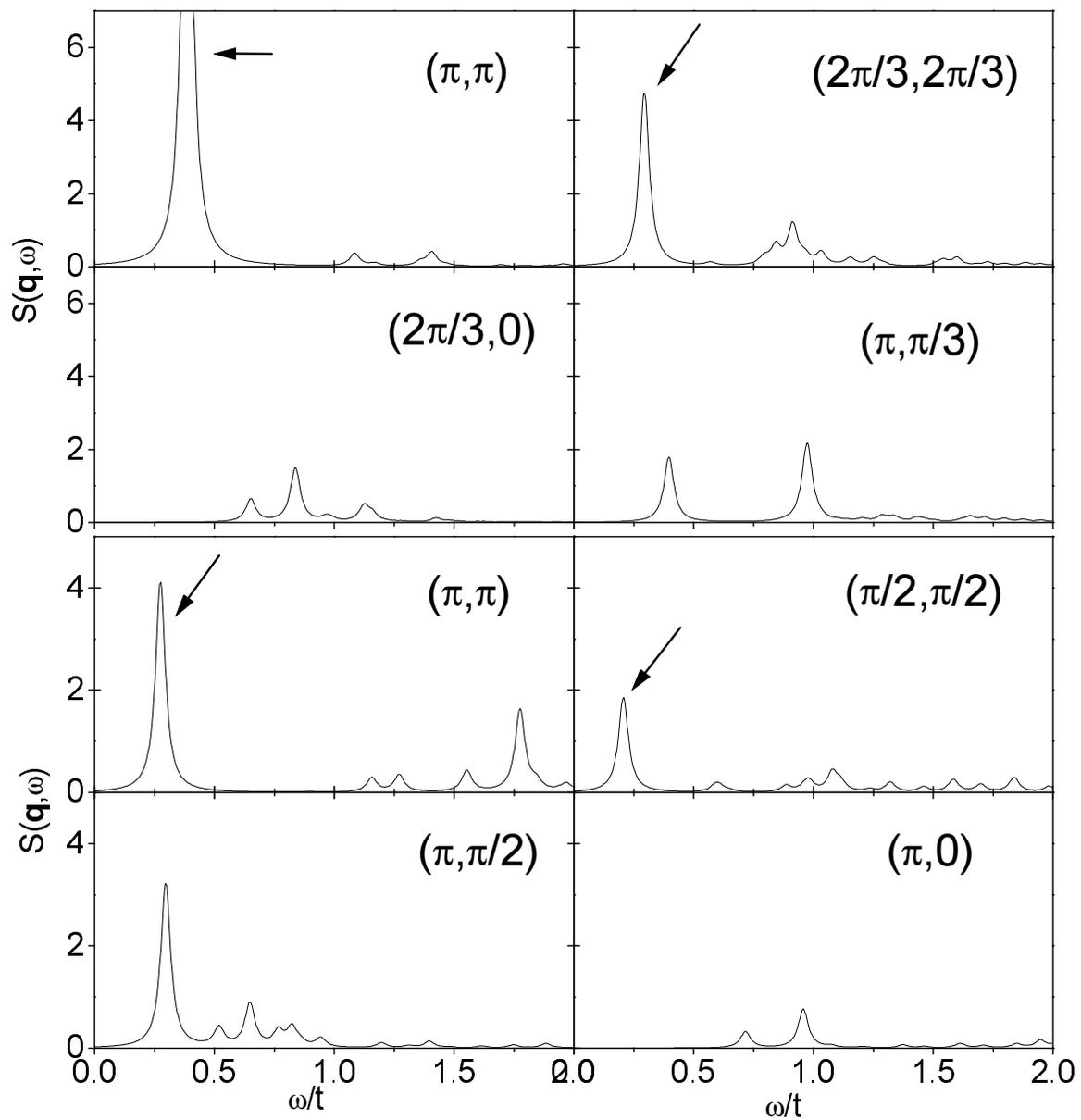
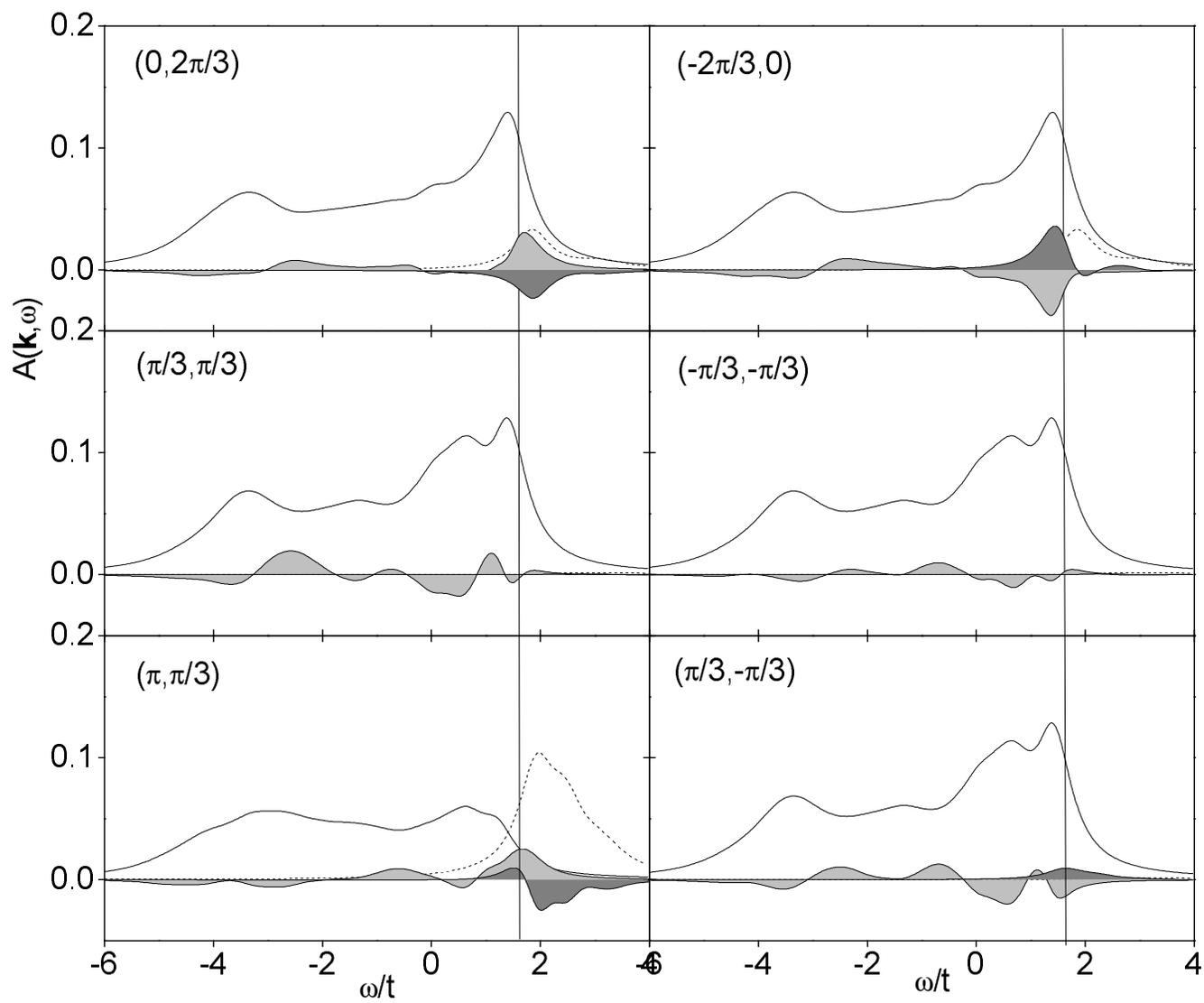
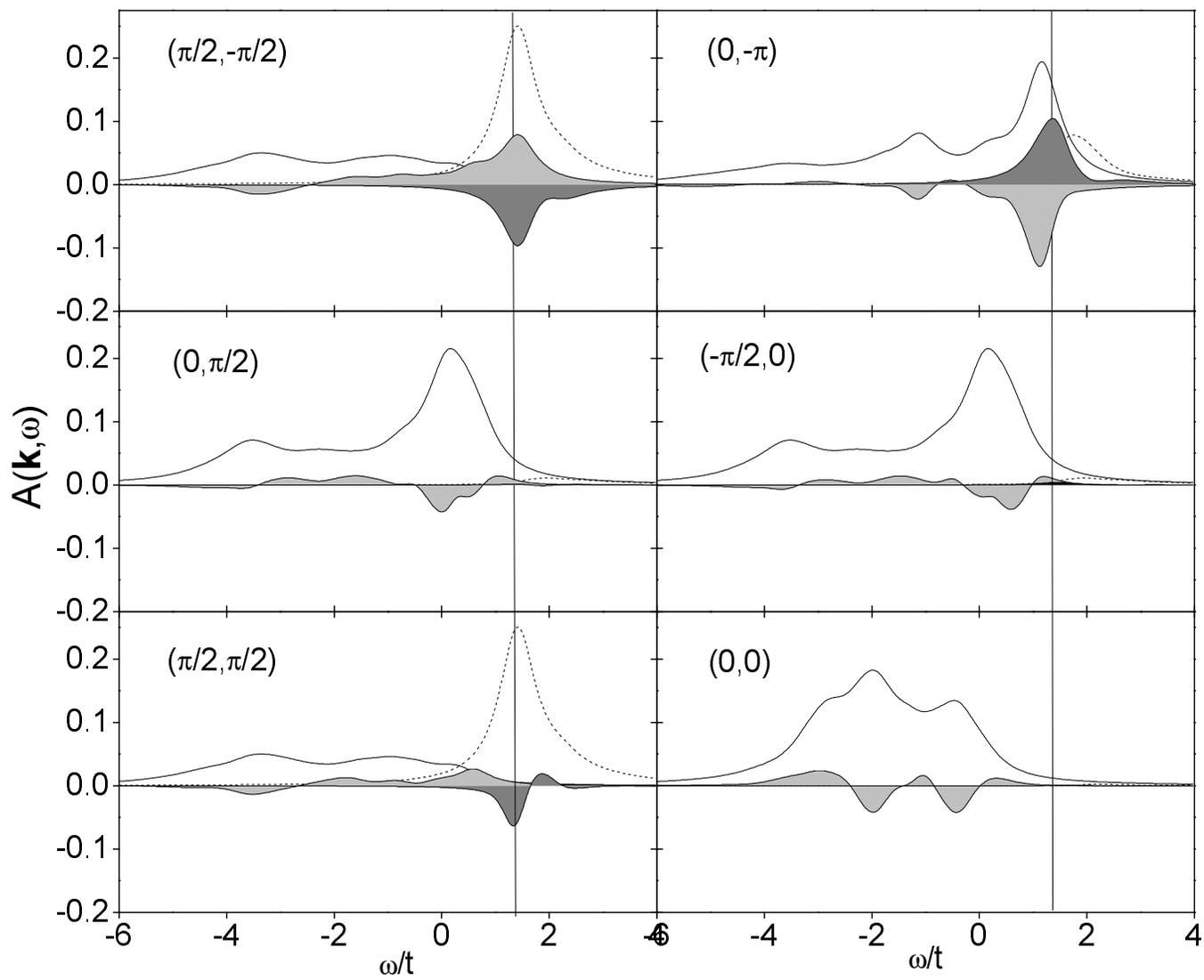
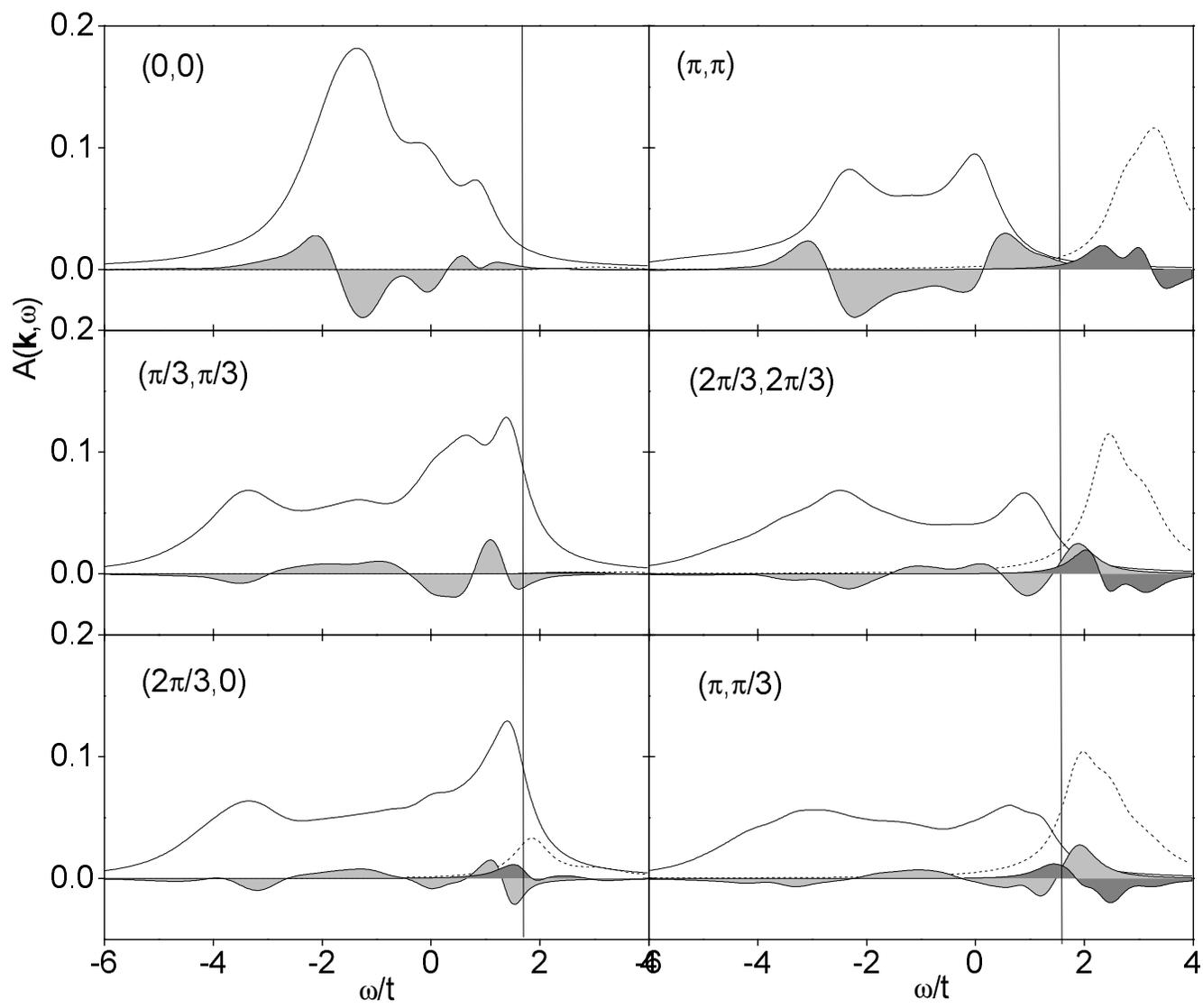
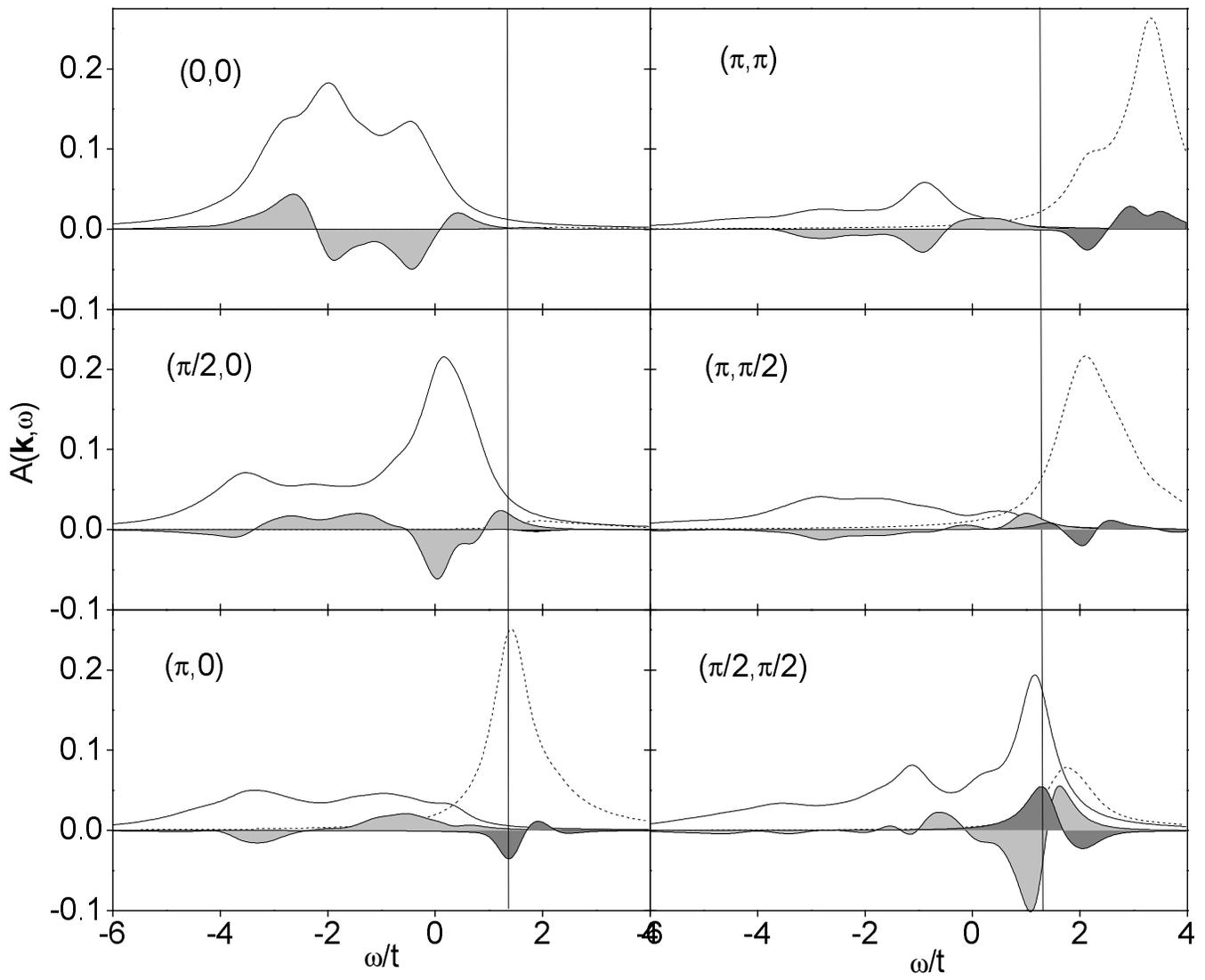
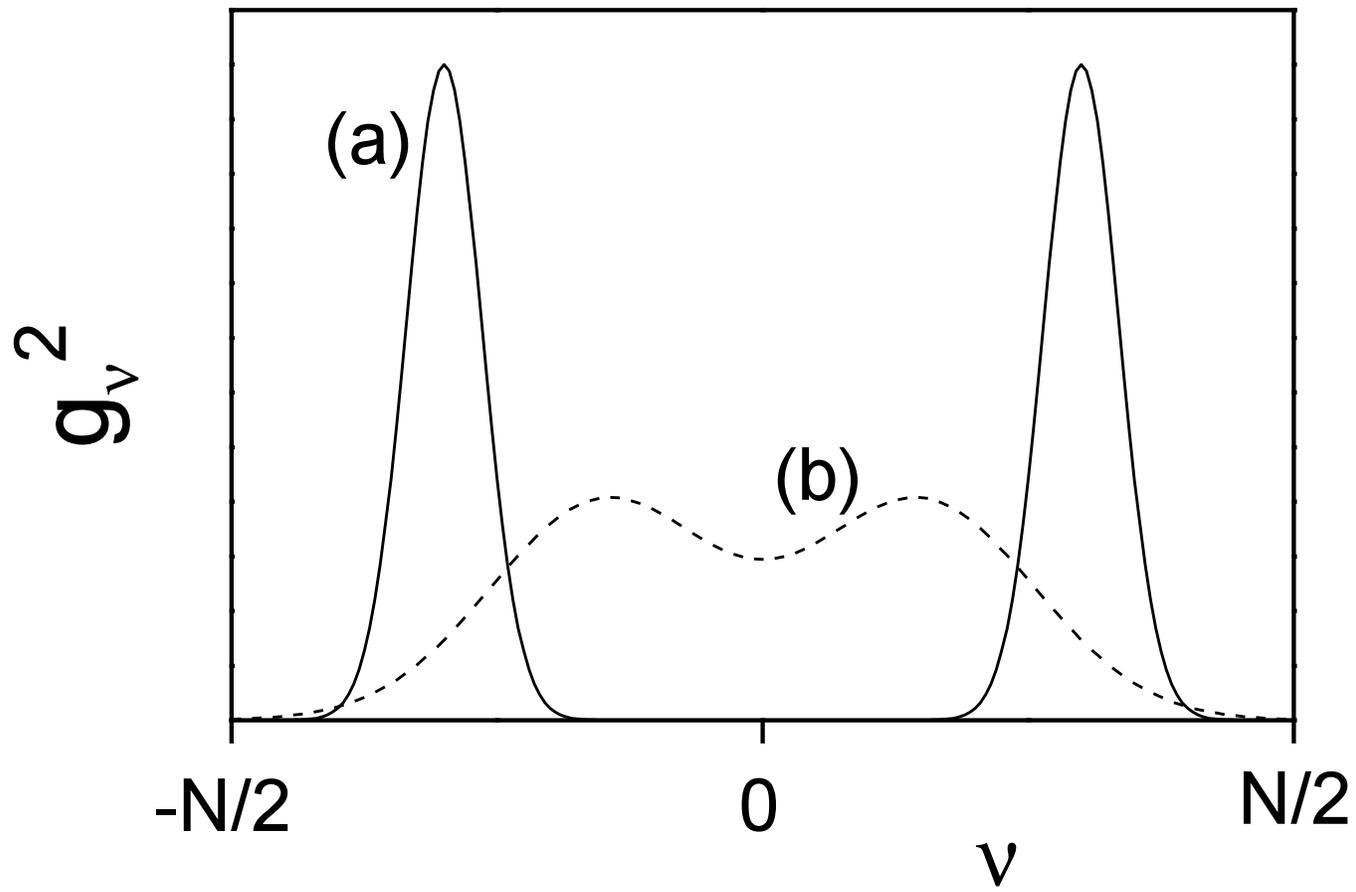
\end{document}